# The Evolution of Investor Activism in Japan

Ryo Sakai
[ryo.sakai@vanderbilt.edu]

May 2, 2022

**Abstract**

Activist investors have gradually become a catalyst for change in Japanese companies. This study examines the impact of activist board representation on firm performance in Japan. I focus on the only two Japanese companies with activist board representation: Kawasaki Kisen Kaisha, Ltd. ("Kawasaki") and Olympus Corporation ("Olympus"). Overall, I document significant benefits from the decision to engage with activists at these companies. The target companies experience greater short- and long-term abnormal stock returns following the activist engagement. Moreover, I show operational improvements as measured by return on assets and return on equity. Activist board members also associate with important changes in payout policy that help explain the positive stock returns. My findings support the notion that Japanese companies should consider engagements with activist investors to transform and improve their businesses. Such interactions can lead to innovative and forward-thinking policies that create value for Japanese businesses and their stakeholders.

*"What is wrong with making money?"* Yoshiaki Murakami (A Japanese activist investor)
(Quote from his June 5, 2006 press conference.)

**Introduction**

Since Yoshiaki Murakami said this word in 2006, Japanese companies have debated whether activist funds are their enemies or allies. After the collapse of the real estate and stock market bubble in the late 1980s, foreign investors actively played important roles in the Japanese market. Amid the emergence of foreign investors in the 1990s, US-style activist funds were introduced to the Japanese market and gradually were recognized by Japanese companies. As Gillan and Starks (1998) defined in their research, activist funds are large shareholders who try to change the status quo through voice without a change in control of the firm. Activist funds have significantly affected Japanese companies in many ways. I analyze these effects in this study.

According to previous research (Miyauchi & Takeda, 2021), two waves of activist funds have existed since the activist funds were introduced to the Japanese market. In the first wave from 2000 to 2008, the main players were foreign and domestic *hostile* activist funds such as Steel Partners and M&A Consulting, well known as the "Murakami Fund". They targeted small and lesser-known listed companies that had underperformed for a long time but retained tremendous amounts of cash and cash equivalents. After acquiring certain amounts of the target companies' stocks, these funds generally requested drastic changes in the target companies. Hamano and Matos (2018) described that the results in the target companies from the first wave were positive stock returns and increases in payout policy (e.g., dividends); however, no significant operational performance improvements in the target companies were observed.

In contrast, in the second wave after 2013, when the Japanese Revitalization Strategy was approved by the Japanese cabinet office, the main players were replaced with foreign and domestic *engagement-oriented* activist funds such as Taiyo Pacific Partners. Unlike the main players during the first wave, engagement-oriented activists did not take hostile actions but rather focused on dialogues with the target companies. In addition, they did not target underperforming and cash-rich listed companies but started expanding their target pools to vertically integrated companies and those with significant off-balance sheet assets (Miyauchi & Takeda, 2021). Like the results of the first wave, increases in payouts in the target companies were observed in the second wave (Miyauchi & Takeda, 2021). In addition, only limited operational performance improvements, such as higher return on assets (ROA), were observed (Miyauchi & Takeda, 2021).

Japanese companies initially reacted unfavorably in response to the emergence of activist funds. Target companies reluctantly accepted only a part of the changes advocated by activist funds. Furthermore, Japanese courts also showed unfavorable reactions to activist funds. Moreover, Japanese institutional investors were reluctant to support proposals from activist funds (Becht, Franks, Grant & Wagner, 2017). However, as time passed, the general views of activist funds are changing. For example, Japanese society started having positive views of activist funds in the second wave (Hamano and Matos, 2018). This view likely emerged as investors and courts realized that activists can help maximize shareholder value in Japanese companies.

To provide insights into how activist funds affect Japanese companies, I deeply study two unique Japanese companies: Kawasaki Kisen Kaisha, Ltd. (Kawasaki) and Olympus Corporation (Olympus). These are the only two companies that have accepted a board member from an activist fund among the approximate 3,000 companies listed in Japan. By studying Kawasaki and Olympus, I shed light on how board members from activist funds affect Japanese companies and whether any differences and similarities in these relations exist versus previous studies.

**Hypothesis**

I propose three hypotheses. The first hypothesis is that Kawasaki and Olympus had positive stock returns both in the short term and in the long term after they accepted a board member from activist funds. The second hypothesis is that after they accepted activist board members, their performance measurements, return on equity (ROE) and ROA, improved significantly because of the activist funds' engagements. The third hypothesis is that these two companies changed their payout policies by increasing either dividends or stock buybacks. These three hypotheses are studied using the methodologies described below.

**Methodology**

To analyze the short-term stock returns of the two companies after the activist engagement, an event study methodology is applied. The event day in this study is the date when the engagement was publicly disclosed, and Kawasaki and Olympus submitted disclosure reports to the Tokyo Stock Exchange. The capital asset pricing model (CAPM) is employed for the estimation of normal returns. Regarding the estimation window, a period between 200 days before the event day and 21 days before the event day is defined. The event period is between 20 days before the event day and 20 days after the event day. Abnormal returns are calculated as the difference between the actual returns and the estimated normal returns based on the CAPM. These abnormal returns are summed to calculate cumulative abnormal returns (CARs) for the event period. Finally, I conduct a T-test to determine if the CARs are statistically different from zero.

For the analysis of the long-term stock returns, the buy and hold abnormal return (BHAR) methodology is employed. It is assumed that a person bought a stock of the two companies on the event day at a market-close price. Then, the person kept the stock for the following one year and two years. Finally, the person sold the stocks at end of each period. I use the buy and hold returns of the Tokyo Stock Price Index (TOPIX) index as the market benchmark. The differences between the TOPIX and the returns of the two stocks constitute the BHAR. For comparison, the BHARs are annualized.

To analyze whether target companies experience improvements in ROA and ROE, I use the Chow test methodology. The Chow test, alternatively called the F-test for structural change, is used to analyze whether there are any structural changes before and after a particular event. The pre-announcement period for Kawasaki is set between the Japanese fiscal fourth quarter of 2011 (March 2012) and the Japanese fiscal fourth quarter of 2018 (March 2019) because the event was announced in April 2019. The post-announcement period of Kawasaki is set between the Japanese fiscal first quarter of 2019 (June 2019) and the Japanese fiscal third quarter of 2021 (December 2021). The pre-announcement period for Olympus is set between the Japanese fiscal fourth quarter of 2011 (March 2012) and the Japanese fiscal third quarter of 2018 (December 2018) because the event was announced in January 2019. The post-announcement period of Olympus is set between the Japanese fiscal fourth quarter of 2018 (March 2019) and the Japanese fiscal third quarter of 2021 (December 2021). Then, the null hypothesis that the pre-announcement is equal to the post-announcement in the ROA and ROE of Kawasaki and Olympus is tested.

Regarding the third hypothesis, the annual financial data related to payouts such as the amounts of dividends and stock buybacks since the fiscal year 2012 are depicted in figures and analyzed below based on these patterns.

**Data**

The financial data of each company is retrieved from the Refinitiv Eikon. Some financial data of Olympus are missing in the Refinitiv Eikon due to revisions following instances of accounting frauds in 2011. I extract the missing data directly from the company's disclosed financial statements, which include restated financials after the frauds were revealed.

**Summary of Analysis**

The event study methodology shows that both Kawasaki and Olympus have positive CARs during the event period. Both CARs are statistically significant, indicating the returns are non-negative. However, the slope of each CAR has a different shape. In addition, while Olympus' CARs remain high after the event, the CARs of Kawasaki gradually decline for 20 days after the event. While Kawasaki shows a negative annualized stock return in the first year in the buy and hold analysis, in the second year it turns out to be a large positive. Olympus shows positive annualized stock returns in the first year and second year in the buy and hold analyses. Given these facts, the data support the first hypothesis.

The results of the Chow test show that Kawasaki has a structural change after the announcement of the acceptance of the board member from the fund. In contrast, it seems that Olympus' ROE and ROA are not statistically significantly affected by the event. However, Kawasaki and Olympus have overall positive slope curves of regressions after the event, indicating that they operated well with their activist fund board member. To sum up, the results support my second hypothesis.

Regarding changes in payouts of the two companies, there are interesting insights. As opposed to the previous research, Kawasaki terminated dividends and stock buybacks after it accepted the activist fund board member. Contrarily, Olympus significantly increased its payouts to shareholders through its stock buybacks. In conclusion, for Kawasaki, the data reject the third hypothesis. However, for Olympus, the results support my third hypothesis.

**Overview of Kawasaki and Olympus**

Kawasaki is a Japanese company operating mainly in the marine transportation business. It was established in 1919, when World War I stimulated huge demands for ships. Kawasaki has three main business segments. The dry bulk segment oversees the transportation by dry bulkers of raw materials such as iron ore, cereal crops, and paper. The energy resource segment includes businesses related to energy such as the transportation of oil, the transportation of natural gas, and the development of energy resources. The product logistics segment manages the transportation of cars and general logistics. In addition, Kawasaki runs a ship management business, a travel agency business, and a real estate leasing and management business. Although Kawasaki ranked third place in the marine transportation industry in Japan for a long time, it experienced significant, consecutive decreases in sales before the board of directors from the activist fund joined in 2019. Furthermore, the global competition in the industry is fierce. Therefore, even other top Japanese marine transportation companies, Nippon Yusen Kaisha and Mitsui O.S.K Lines, Ltd, have received high pressure to change from their shareholders.

**Figure 1**     Kawasaki's stock prices and TOPIX during the study period

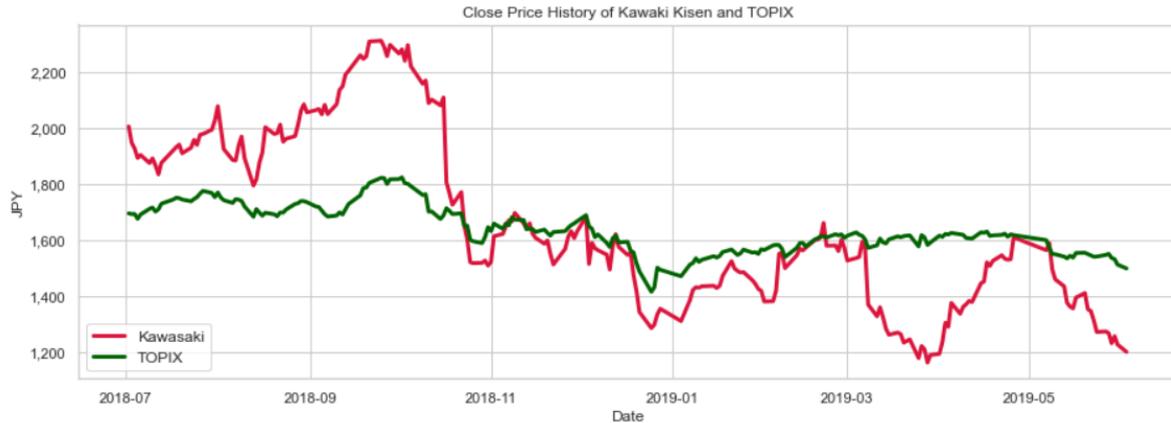

Olympus is a Japanese company operating in the manufacture and sale of precision machinery and instruments. It was also established in 1919 as a manufacturer of microscopes. Olympus has four business segments. The endoscopy segment provides services related to medical endoscopy. The therapeutic solutions segment oversees the manufacture and sales of innovative surgical therapeutical devices. The scientific solutions segment operates business in research devices such as biological microscopes and X-ray fluorescence analyzers. The other business segment develops and sells biomaterials and orthopedic devices. In 2011, it was revealed that Olympus had conducted several accounting frauds for a long time. As a result, Olympus was almost delisted. However, the new management team revamped Olympus from the edge of death although the company's reputation was seriously damaged. Because the management of the company was completely replaced after the fraud, it is said that Olympus has changed its corporate culture from the traditional Japanese company culture to a more flexible and forward-thinking company culture.

**Figure 2**     Olympus's stock prices and TOPIX during the study period

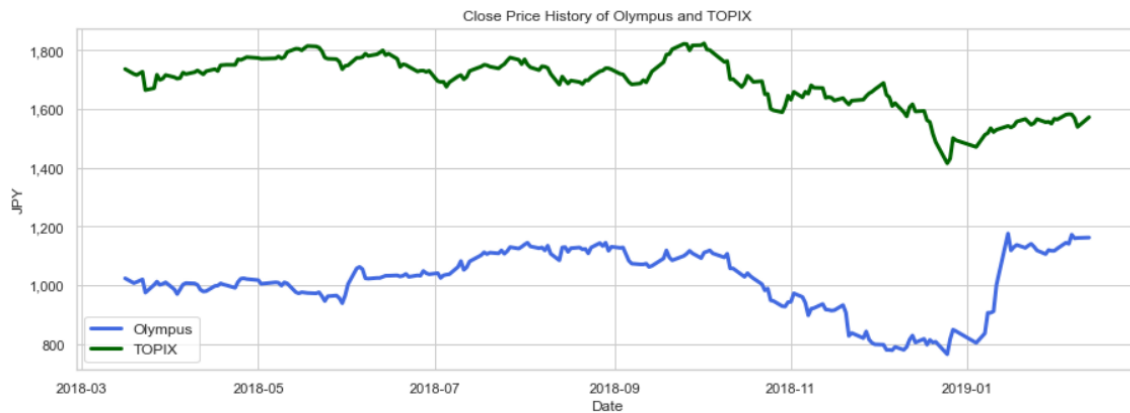

**How to React to Activist Funds**

Effissimo Capital Management Pte Ltd (Effissimo), an activist fund based in Singapore but run by Japanese professionals, disclosed that it acquired 6.18% of Kawasaki's stocks on August 31, 2015. After the first announcement, Effissimo used the creeping takeover strategy, which is a takeover that involves gradually buying stocks of a company from different shareholders until a fund has enough to take control of the company (Creeping takeover, n.a.). As a result of the creeping takeover strategy, in June 2018, Effissimo had 38.99% of the company's stocks. During the three years from 2015, Effissimo kept pressuring Kawasaki to change through proactive actions such as voting down, at the shareholders' meeting in 2016, the proposal from Kawasaki for the appointment of the CEO, Mr. Eizo Murakami.

In response to Effissimo's buying of the stocks, Kawasaki did not show a welcome attitude toward Effissimo in 2015. However, after the rejection of the appointment proposal in 2016, Kawasaki recognized the threat from Effissimo and gradually changed its attitude toward Effissimo, trying to communicate with Effissimo. According to Nikkei's news article on June 23, 2017, the management of Kawasaki met with a representative director of Effissimo, Mr. Takushi Takasaka, and directly explained its mid-term corporate strategy to him. By that time, Kawasaki seems to have already given up preventing Effissimo from buying its stocks. Finally, on April 26, 2019, Kawasaki announced that Mr. Ryuhei Uchiyama from Effissimo would be nominated as a candidate for a board member at the meeting of shareholders in 2019. On June 21, 2019, Mr. Uchiyama was officially appointed to the board of directors of Kawasaki.

ValueAct Capital Management LP (ValueAct), a San Francisco-based activist fund, started acquiring less than 5%, which is the threshold of the large shareholding report in Japan, of Olympus stocks in 2016. (Reuters, 2021). On May 25, 2018, ValueAct reported officially that it had 5.03% of Olympus stocks. Olympus is the first Japanese company for ValueAct to invest in. Unlike Effissimo, ValueAct kept holding almost the same share of Olympus stocks after the first disclosure. In addition, ValueAct did not too much pressure on Olympus but rather tried to engage in dialogues with Olympus.

Unlike Kawasaki, Olympus took different approaches to ValueAct. From the beginning, Olympus reacted favorably to ValueAct. After 2016, the management team in Olympus kept actively communicating with ValueAct (Reuters, 2021). Furthermore, Olympus tried to use the power of people from outside to accelerate its challenge to be a global firm when it received an initial proposal from ValueAct, the proposal that says that Mr. Robert Hale and Mr. Jimmy Beasley should be board candidates (Matsuyama, 2022). Eventually, on January 11, 2019, Olympus announced that Mr. Hale would be nominated as a candidate for a board member. Mr. Hale was officially appointed to the board of directors on June 25, 2019.

**Details of Analysis**

*Event Study*

The event study is an empirical analysis revealing how a stock is likely to react to a particular event. More specifically, using a particular market benchmark, normal returns are estimated. Then, comparing the estimated normal returns with actual returns, abnormal returns and CARs are calculated. Finally, the T-test is conducted to test the null hypothesis ($H_0$: the CARs = 0). In the following parts, each process will be explained.

*(1) Calculation of Estimated Normal Returns from the CAPM*

I assume that normal returns would have been realized if the event did not occur. In this study, the CAPM is employed to calculate normal returns. The CAPM is defined as below. As the market benchmark, the TOPIX is used. In addition, because long-term yields of Japanese government bonds were either very close to 0 percent or negative during the study period, the risk-free rate is assumed as 0.

$$R_{i,t} = \alpha_i + \beta_i R_{m,t} + \varepsilon_{i,t}$$

$R_{i,t}$: The return of stock $i$ for a certaion time of $t$
$R_{m,t}$: The return of Nikkei 225 for a certaion time of $t$
$\alpha_i, \beta_i$: Stock specific constants
$\varepsilon_{i,t}$: Prediction error for stock $i$

Using the ordinary least squares regression (OLS) method, $\hat{\alpha}_i$ and $\hat{\beta}_i$ are estimated during the window period from 200 days before the event day to 21 days before the event day. As a result of the OLS method, abnormal returns are estimated as an equation, $\hat{R}_{i,t} = \hat{\alpha}_i + \hat{\beta}_i R_{m,t}$.

The below equations show ones for Kawasaki and Olympus. In addition, Table 1 and 2 illustrate some key numbers of the OLS.

$$R_{Kawasaki,t} = -0.2199 + 1.2541\, R_{TOPIX,t}$$

**Table 1**     Regression results of the CAPM for Kawasaki

OLS Regression Results

| | | | |
|---|---|---|---|
| Dep. Variable: | y | R-squared: | 0.257 |
| Model: | OLS | Adj. R-squared: | 0.253 |
| Method: | Least Squares | F-statistic: | 61.49 |
| Date: | Mon, 25 Apr 2022 | Prob (F-statistic): | 3.94e-13 |
| Time: | 22:19:51 | Log-Likelihood: | -420.79 |
| No. Observations: | 180 | AIC: | 845.6 |
| Df Residuals: | 178 | BIC: | 852.0 |
| Df Model: | 1 | | |
| Covariance Type: | nonrobust | | |

| | coef | std err | t | P>\|t\| | [0.025 | 0.975] |
|---|---|---|---|---|---|---|
| const | -0.2199 | 0.188 | -1.170 | 0.243 | -0.591 | 0.151 |
| x1 | 1.2541 | 0.160 | 7.841 | 0.000 | 0.938 | 1.570 |

| | | | |
|---|---|---|---|
| Omnibus: | 87.211 | Durbin-Watson: | 1.924 |
| Prob(Omnibus): | 0.000 | Jarque-Bera (JB): | 855.748 |
| Skew: | -1.516 | Prob(JB): | 1.50e-186 |
| Kurtosis: | 13.242 | Cond. No. | 1.18 |

$$R_{Olympus,t} = -0.1022 + 0.9544\, R_{TOPIX,t}$$

**Table 2**     Regression results of the CAPM for Olympus

| | OLS Regression Results | | |
|---|---|---|---|
| Dep. Variable: | y | R-squared: | 0.296 |
| Model: | OLS | Adj. R-squared: | 0.292 |
| Method: | Least Squares | F-statistic: | 74.90 |
| Date: | Mon, 25 Apr 2022 | Prob (F-statistic): | 2.88e-15 |
| Time: | 22:19:51 | Log-Likelihood: | -316.02 |
| No. Observations: | 180 | AIC: | 636.0 |
| Df Residuals: | 178 | BIC: | 642.4 |
| Df Model: | 1 | | |
| Covariance Type: | nonrobust | | |

| | coef | std err | t | P>\|t\| | [0.025 | 0.975] |
|---|---|---|---|---|---|---|
| const | -0.1022 | 0.105 | -0.973 | 0.332 | -0.309 | 0.105 |
| x1 | 0.9544 | 0.110 | 8.655 | 0.000 | 0.737 | 1.172 |

| | | | |
|---|---|---|---|
| Omnibus: | 58.354 | Durbin-Watson: | 1.793 |
| Prob(Omnibus): | 0.000 | Jarque-Bera (JB): | 283.046 |
| Skew: | -1.114 | Prob(JB): | 3.45e-62 |
| Kurtosis: | 8.725 | Cond. No. | 1.06 |

*(2) Calculation of Cumulative Abnormal Returns*

First, abnormal returns are calculated by subtracting the estimated normal returns and actual returns as below.

$$AR_{i,t} = R_{i,t} - \hat{R}_{i,t}$$

Then, CARs between 20 days before the event day to 20 days after the event day are calculated. Figure 3 shows the CARs of Kawasaki and Olympus.

$$CAR_{(-20,20)} = \sum_{t=-20}^{20} AR_{i,t}$$

**Figure 3**     Cumulative abnormal returns of Kawasaki and Olympus for the event period

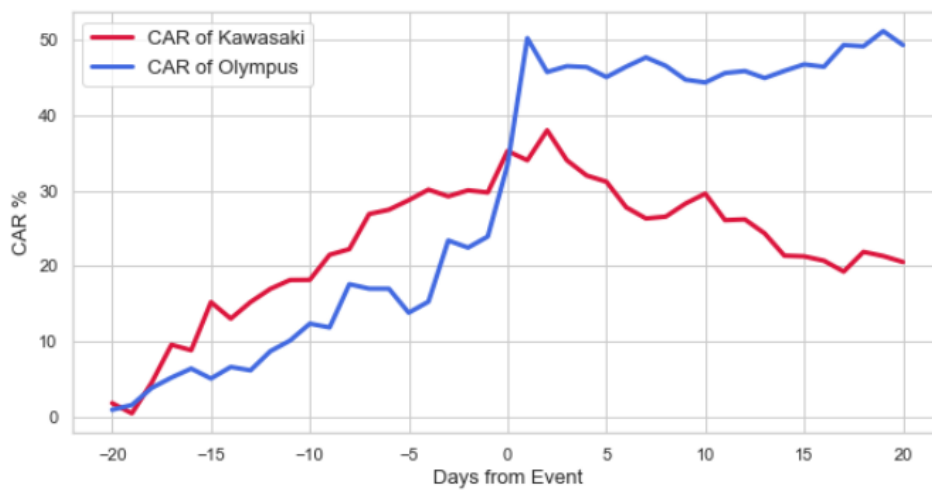

*(3) T-test for the Cumulative Abnormal Returns*

To test whether the CARs observed are statistically significantly different from negative numbers, the T-test for each of the CARs is conducted. Because I would like to understand whether the cumulative abnormal return is positive, I use a one-tailed test is used. In addition, 5% is set as the level of significance for the test.

$H_0$: The cumulative abnormal return $\leq 0$
$H_1$: The cumulative abnormal return $> 0$

As Table 3 shows, the null hypothesis should be rejected because both p-values are smaller than 0.025. Thus, positive abnormal returns existed in the event period.

Table 3     Results of the T-test for the CARs of Kawasaki and Olympus

|  | T_value | P_value |  | T_value | P_value |
|---|---|---|---|---|---|
| Kasaki CAR | 16.38700 | 0.00000 | Olympus CAR | 10.17424 | 0.00000 |

*Buy and Hold Analysis*

The buy and hold analysis assumes that an investor buys a stock on the event day and holds the position for a long time. Then, returns during the holding period are calculated. Following the total stock return equation, one-year returns and two-year returns for Kawasaki, Olympus, and TOPIX are calculated.

$$Total\ Stock\ Return = \sqrt[h]{\frac{(Stock\ Price_t - Stock\ Price_{t+h}) + Dividends_h}{Stock\ Price_t}} - 1$$

$t$: The event date

$h$: The holding peirod

As Table 4 and Figure 4 show the numbers, Olympus has 60.10% as the one-year BHAR and 34.86% as the annualized two-year BHAR. While Kawasaki has a negative one-year BHAR, −24.47%, it has a positive two-year BHAR of 23.90%.

Table 4     Summary of Kawasaki's and Olympus's BHAR calculations

|  | sp_start_close | sp_1yrend_close | sp_2yrend_close | topix_start_close | topix_1yrend_close | topix_2yrend_close | dividend_1yr | dividend_2yr | ar_sp_1yr | ar_sp_2yr | ar_topix_1yr | ar_topix_2yr | bhr_1yr | bhr_2yr |
|---|---|---|---|---|---|---|---|---|---|---|---|---|---|---|
| Kawasaki | 1610 | 1048 | 2835 | 1,617.93 | 1,449.15 | 1,914.98 | 0 | 0 | -34.91 | 32.70 | -10.43 | 8.79 | -24.47 | 23.90 |
| Olympus | 1,001.25 | 1,733.50 | 2,097.00 | 1,529.73 | 1,740.53 | 1,854.94 | 7.50 | 17.50 | 73.88 | 44.98 | 13.78 | 10.12 | 60.10 | 34.86 |

**Figure 4**     Kawasaki's and Olympus's Buy&Hold Abnormal Return

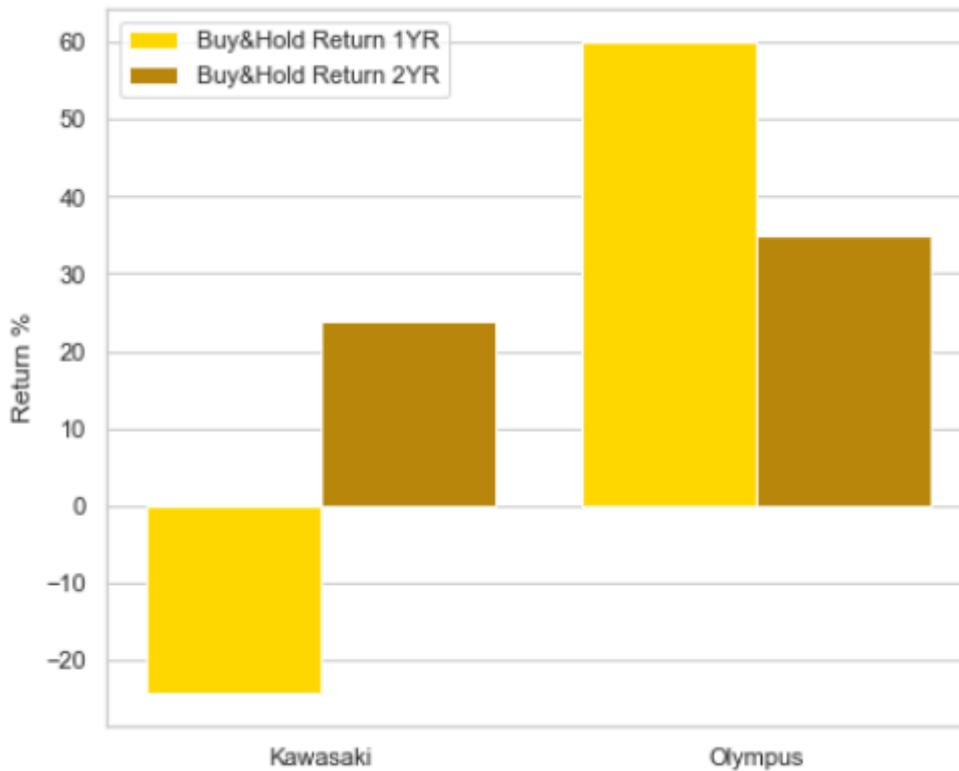

*Chow Test*

The Chow test, alternatively called the F-test for structural change, is used to analyze whether there are any structural changes before and after a particular event. In this study, the ROA and ROE of Kawasaki and Olympus are tested by the Chow test. In addition, quarters before the event are defined as the prior period while quarters after the event are defined as the latter period. Then, the null hypothesis is tested with a 5% level of significance.

$H_0$: The coefficient of regression for the prior period = that of regression for the latter period
$H_1$: Otherwise

*(1) ROA of Kawasaki*
 ① The entire period (2012/3 ~ 2021/12)
  Y = – 2.069 + 0.131 X
  $R^2$ = 0.143
  SSR = 543.269
 ② The prior period (2012/3 ~ 2019/3)
  Y = 1.105 – 0.125 X
  $R^2$ = 0.189
  SSR1 = 137.120
 ③ The latter period (2019/6 ~2021/12)
  Y = – 42.825 + 1.345 X
  R2 = 0.756
  SSR2 = 64.154
 ④ F-value

F-value = 30.585

Under the 5% level of significance, the critical value of the F distribution with a degree of freedom of 2 (denominator) and 36 (numerator) is about 3.26. Therefore, the null hypothesis is rejected. Thus, the ROA of Kawasaki appears to be affected by the activist event.

**Figure 5**     Chow test result for Kawasaki's ROA

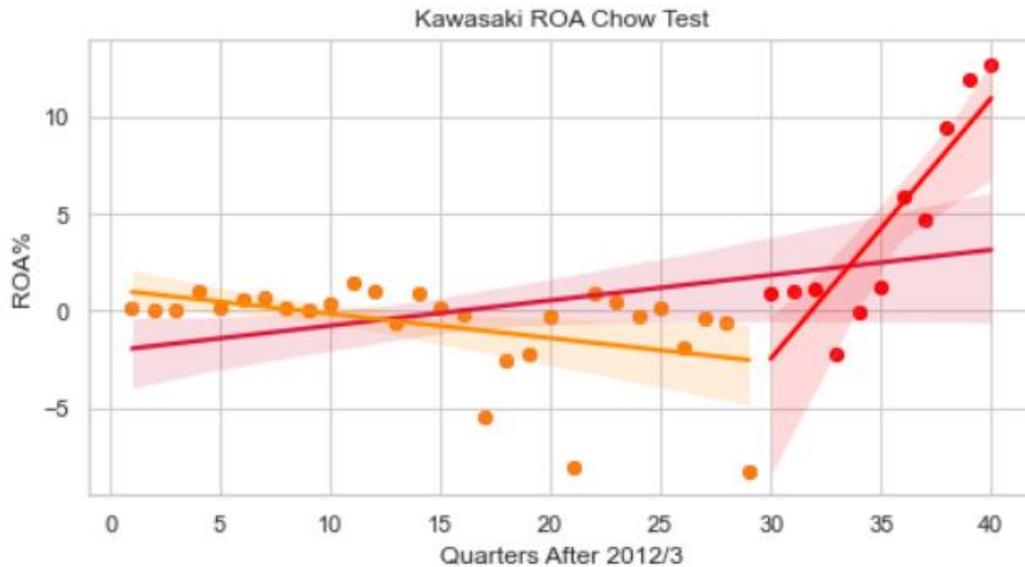

*(2) ROE of Kawasaki*

 ① The entire period (2012/3 ~ 2021/12)
  Y = – 5.060 + 0.274 X
  $R^2$ = 0.068
  SSR = 5512.829

 ② The prior period (2012/3 ~ 2021/3)
  Y = – 4.606 – 0.525 X
  $R^2$ = 0.186
  SSR1 = 2444.397

 ③ The latter period (2019/6 ~2021/12)
  Y = – 88.283 + 2.826 X
  R2 = 0.639
  SSR2 = 495.636

 ④ F value
  F value = 15.752

Under the 5% level of significance, the critical value of the F distribution with a degree of freedom of 2 (denominator) and 36 (numerator) is about 3.26. Therefore, the null hypothesis should be rejected. The ROE of Kawasaki appears to be affected by the activist event.

**Figure 6**   Chow test result for Kawasaki's ROE

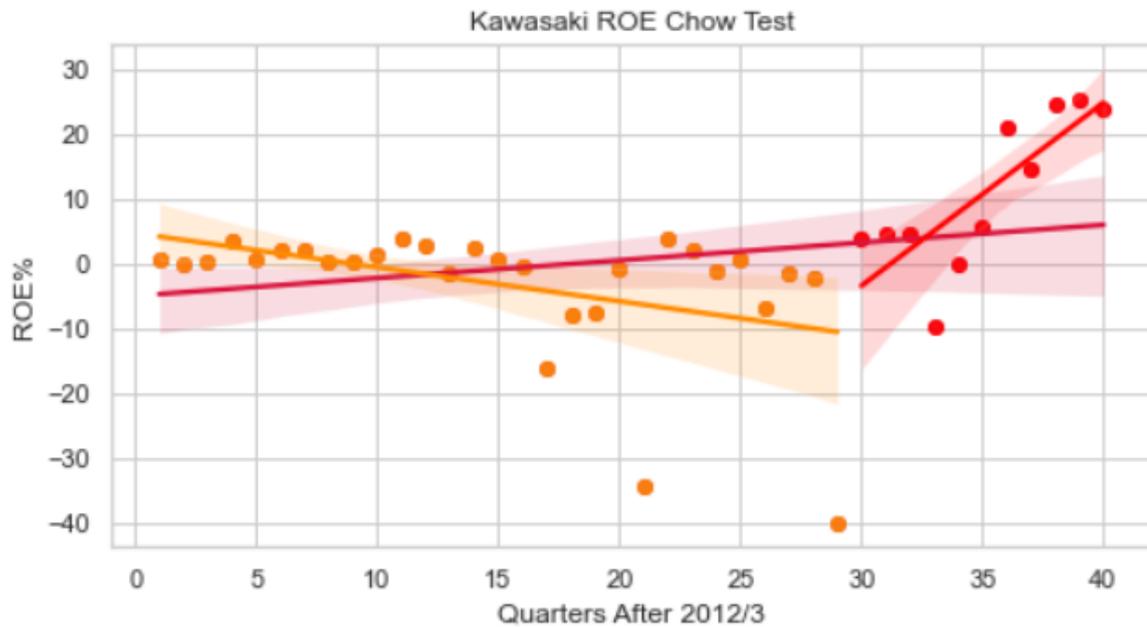

*(3) ROA of Olympus*
- ① The entire period (2012/3 ~ 2021/12)
  Y = − 0.116 + 0.051 X
  $R^2$ = 0.209
  SSR = 52.685
- ② The prior period (2012/3 ~ 2018/12)
  Y = − 0.088 + 0.050 X
  $R^2$ = 0.104
  SSR1 = 35.788
- ③ The latter period (2019/3 ~2021/12)
  Y = –2.134+ 0.109 X
  R2 = 0.118
  SSR2 = 16.253
- ④ F value
  F value = 0.223

Under the 5% level of significance, the critical value of the F distribution with a degree of freedom of 2 (denominator) and 36 (numerator) is about 3.26. Therefore, the results fail to reject the null hypothesis. In other words, the activist event does not appear to influence the ROA of Olympus.

**Figure 7**    Chow test result for Olympus's ROA

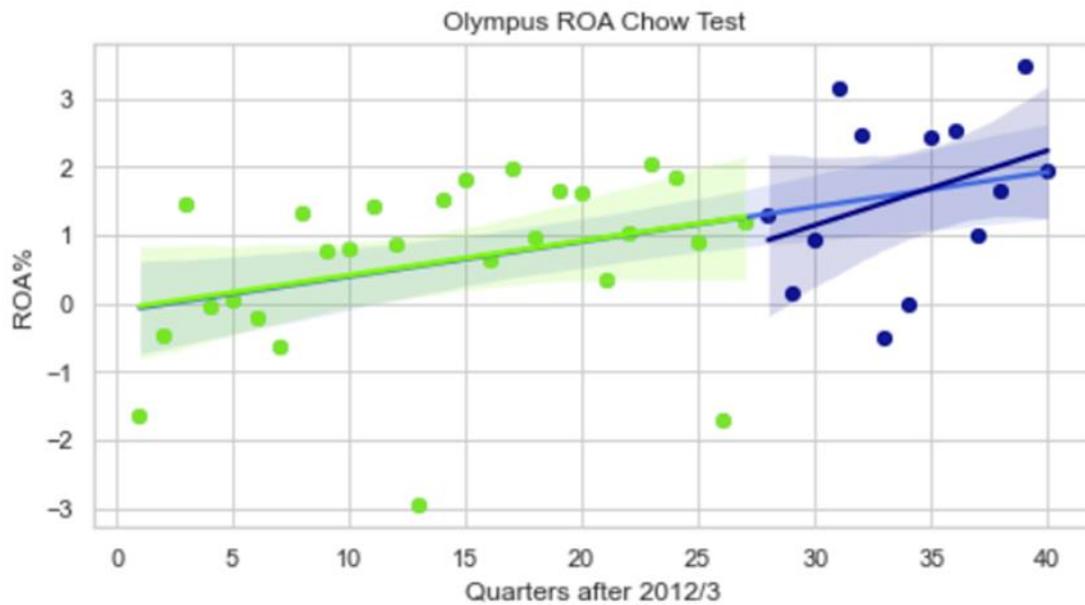

(4) ROE of Olympus
- ① The entire period (2012/3 ~ 2021/12)
  Y = – 2.607 + 0.220 X
  $R^2$ = 0.107
  SSR = 2152.267
- ② The prior period (2012/3 ~ 2018/12)
  Y = –3.254 – 0.284 X
  $R^2$ = 0.062
  SSR1 = 2008.950
- ③ The latter period (2019/3 ~2021/12)
  Y = – 8.601 + 0.382 X
  R2 = 0.173
  SSR2 = 126.775
- ④ F value
  F value = 0.139

Under the 5% level of significance, the critical value of the F distribution with a degree of freedom of 2 (denominator) and 36 (numerator) is about 3.26. These results fail to reject the null hypothesis. The activist event does not seem to influence the ROE of Olympus.

**Figure 8**   Chow test result for Olympus's ROE

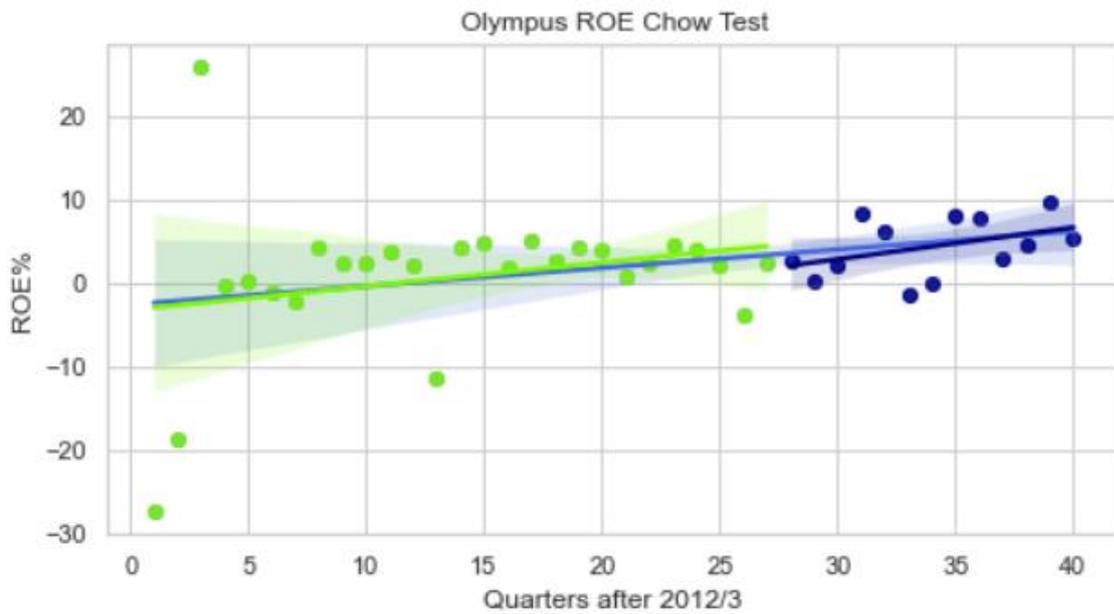

*Payout Ratio*

As shown in the below figures, Kawasaki stopped distributing its retained earnings to shareholders via dividends and stock buybacks after the activist event. However, Olympus kept paying its dividends after Mr. Hale became on the board of directors. Moreover, it also increased the amounts of stock buybacks.

**Figure 9**   Other financial metrics in Kawasaki

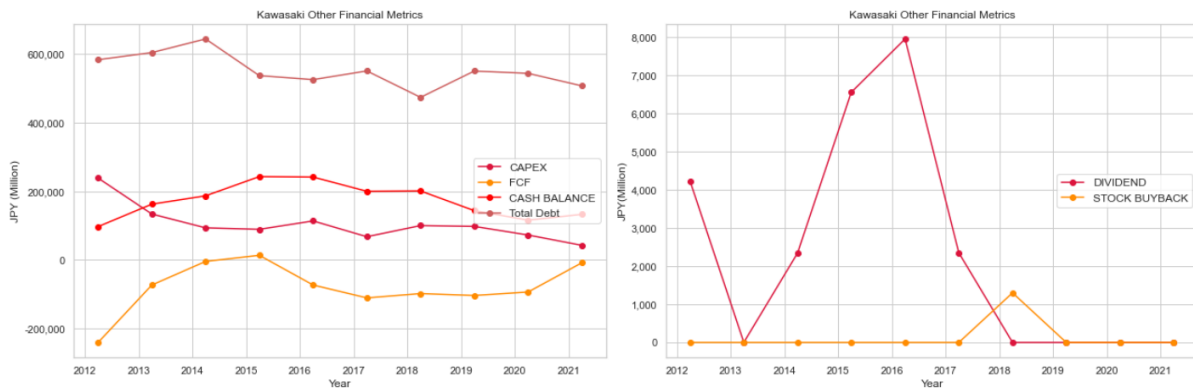

**Figure 10**   Other financial metrics in Olympus

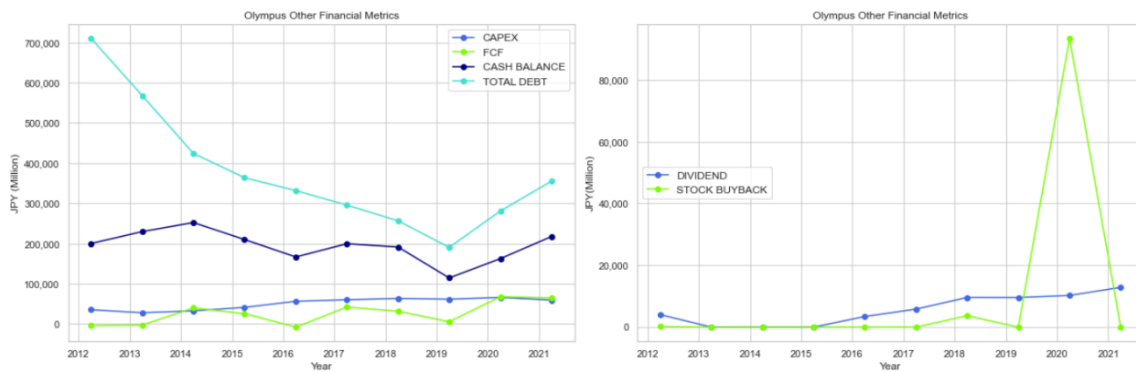

**Interpretation of Results**

*Event Study*

The positive CARs are observed in Figure 3 even before the events were announced. The phenomena imply two possible explanations. One possibility is that some investors might have obtained some insider information about the announcements. However, given the fact that restrictions on insider trades were strict in those days, it could be more reasonable to think that some rational investors must have expected Kawasaki and Olympus to accept soon a board member from ValueAct and Effissimo respectively.

Although both CARs of Kawasaki and Olympus during the event period show a statistically significant difference from negative values, there are differences in the magnitude and slope of returns. In Figure 3, Olympus has a big spike of CARs around the event day, while the CARs of Kawasaki increase gradually and do not experience any sudden spikes. This differential response could be explained by the difference in market participants' expectations. Because Olympus is the first Japanese listed company to accept a board member from activist funds, many market participants could have been surprised at the time of the announcement. (As mentioned above, some investors could have expected the event would take place.) In addition, compared with Kawasaki, Olympus reacted to the activist fund positively and kept in conversations with ValueAct even before the event. Therefore, market participants might have anticipated the positive effects of the activist event on Olympus and reacted positively by buying more of Olympus' stock.

While Olympus has higher CARs for 20 days after the event, the CARs at Kawasaki decrease gradually after the event day. This could be because more market participants kept buying Olympus' stock with strong expectations for potential increases in Olympus's immediate paybacks. As Figure 10 shows, Olympus's free cash flows are positive before the event. Furthermore, debts are repaid significantly while its cash positions are stable for a long time. Given these conditions, it is reasonable for investors to have believed that Olympus would start significant paybacks soon. In fact, Olympus borrowed money from the market and increased its stock buybacks after the event.

*Buy and Hold Analysis*

As Figure 4 illustrates, overall Kawasaki and Olympus exceed the returns of the market index, TOPIX. The reason why Kawasaki has only a negative number in the first year could be the spread of COVID19. In 2020, the virus spread hit severely all industries; however, the marine transportation industry was hit much more seriously than other industries. At the beginning of the pandemic, all economic activities were limited. As a result, almost all transportations were shut down, including those at Kawasaki. In other words, the negative returns in 2020 could be due to the pandemic and not the activist event.

*Chow Test*

The results from the Chow Test imply that while Kawasaki operationally improves after the acceptance of a board member from ValueAct, Olympus does not statistically show a change in operational metrics after the event. However, the results need to explore more deeply. After Olympus's accounting frauds in 2011, Olympus transformed itself drastically by 2019. As Figure 10 shows, Olympus's financial metrics such as free cash flows looked healthy as of 2019. I posit that Olympus's management goals must have not been only short-term performance improvements but rather more long-term strategic improvements such as enhancing its corporate governance and establishing good relationships with stakeholders. On the other hand, at the time, Kawasaki was under severe pressure to reform. As Figure 9 illustrates, its free cash flows are almost negative after 2012. In addition, Kawasaki's sales decreased gradually because of the global fierce competition in the marine transportation industry. By accepting Mr. Uchida from Effissimo, Kawasaki used the external force to transform itself for the short time. As a result, it was able to improve its operational performances, thus leading to significant improvements in its ROA and ROE.

*Payout Ratio*

Olympus supported the previous research, increasing payouts through dividends and stock buybacks. In contrast, it seems that Kawasaki showed a different result from the previous studies, stopping dividends and stock buybacks. However, if we look at Kawasaki's situation at the time, this payout policy change makes sense. As mentioned above, it suffered negative free cash flows. In addition, Kawasaki requires significant capital expenditure due to the nature of its business. One way to stop "bleeding cash" is to stop dividend payments to shareholders. However, many companies are hesitant to stop dividend payouts. This is because managers care about negative signals about future cash flows stemming from a payout termination. Generally, many market participants believe that a suspension or cut in dividends reflects companies' knowledge of negative forecasts. However, Kawasaki bravely decided to change its payout policy to improve its cash balance and secure liquidity on hand. Given the fact that Kawasaki did not increase the short-term payouts and did improve its operational performances, the typical criticism toward activist funds that they force companies to increase payout without any strategic thoughts could not be applied to Effissimo.

**Regulations and laws**

Finally, the perspectives from regulations and laws are mentioned to understand the situation surrounding Japanese companies and activist funds in Japan. For a long time, regulations and the Japanese court protected Japanese companies from activist funds. As Kamiya, Kumaki, and Huh (2022) mentioned, Japan's foreign direct investment regime, which includes the Foreign Exchange and Foreign Trade Act (FEFTA) in Japan and other relevant regulations and ordinances, historically has been an impediment to activism in Japan. The law requires foreign investors to follow restrictions such as approval from the Japanese government when they invest in companies that the Japanese government considers important in terms of national security. Nikkei (2020) said that 42% of listed companies in the Japanese Market were protected as of 2020. Most activist funds are foreign funds. As a result, many Japanese companies have avoided engagements with activist funds.

Furthermore, the Japanese court has reacted unfavorably to activist funds as well. As did Steel Partners lose the file about the invalidity of poison pills structured by Bulldog Sauce in 2007, the Japanese court generally has worked against activist funds. In the case of Steel Partners, the Tokyo High Court considered an "abuse bidder" who pursues only its own interest. These judgments stimulated negative images of activist funds among people in Japan.

However, as Okamura (2022) mentioned, since the two codes—the Stewardship code in 2014 and the Corporate Governance in 2015—were issued, Japanese companies have raised their awareness of the importance of corporate governance. Dialogues with activist funds have gradually gained popularity. In addition, more recently, the head of the Tokyo Stock Exchange (TSE), one of the most important market regulators in Japan, urged to start a dialogue with activist funds (Obe, 2022).

To accelerate this momentum, I argue that two points in the current regulations and laws could be changed. First, the current Japanese foreign direct investment regime should be more relaxed. Although the Japanese companies that directly work on national defense or security such as Toyota and Hitachi should be protected, the FEFTA protects now too many companies from potential value-adding activism. For example, the law protects a supermarket operator, a stationery manufacturer, and a spa operator. Such industries do not tie directly to national defense and security. Furthermore, the Ministry of Finance does not disclose the criteria of the protection. Such ineffective and unclear legislation could alienate foreign investors, which include activist funds. The Japanese government should carefully review the way it operates this law and regime.

Second, the reform of the TSE should be accelerated more. Since April 2022, the TSE has introduced three new market categories, Prime, Standard, and Growth. The purpose of the reform is to reduce the number of companies listed in the top tier. Before the reform, 2,184 companies of around 3,000 listed companies were in the previous top tier market, the TSE first. However, the reform has not been fully implemented because the TSE provided some temporal treatments and a grace period to the companies listed in the TSE first. According to the TSE announcement (2022), only 344 companies moved from the TSE first to the Standard, and 1,840 companies remained in the Prime tier. However, some of the 1,840 companies are not qualified as global firms that the TSE wants to join the Prime tier. As the head of the TSE mentioned in an interview, global firms should actively communicate with shareholders, even activist funds. The TSE should push unqualified companies to move to the Standard and recreate the top tier market where shareholders could easily meet companies that are welcome to have dialogues with any shareholders.

**Conclusion**

This paper provides new research on how companies change after accepting a board member from activist funds. Overall, Kawasaki and Olympus received benefits from their decision to engage with activists. Positive short-term abnormal returns and long-term total stock returns were observed. In addition, Kawasaki tremendously improved its ROE and ROA after the event. Although the formal statistical test fails to uncover structural changes in Olympus's operating performance after the event, the regression analysis reveals positive trends in Olympus's ROE and ROA. These facts support the notion that appointing a board member from an activist fund has helped companies to change in a positive way. In addition, my findings support prior academic research in that some target companies, such as Olympus, increase their payouts to investors following an activist event. However, in contrast to prior work, I show that not all companies follow a path of increasing payouts after an activist engagement. For example, Kawasaki did not increase its payouts and instead improved its cash balance and increased the liquidity of the assets on its balance sheet. Then, Kawasaki took a step further, reviewing its business portfolio and reallocating its resources to strengthen its fundamental business capabilities.

In summary, my study highlights the positive impact of activist funds on investor returns and company operations. I argue that Japanese companies and society should not view activist funds as an enemy. Instead, my findings indicate that companies should proactively establish relationships with activist funds as they are a good catalyst for change. I hope that more Japanese companies open their door to activist funds and engage more frequently to accomplish companies' goals.